**José Alberto Benítez-Andrades** [1], **Alejandro Rodríguez-González** [2], **Carmen Benavides** [1], **Leticia Sánchez-Valdeón** [1] and **Isaías García** [3,*]

[1] SALBIS Research Group, Facultad de Ciencias de la Salud, Universidad de León, Campus de Ponferrada Avda/Astorga s/n, C.P. 24402 Ponferrada (León), Spain; jbena@unileon.es (J.A.B.-A.); carmen.benavides@unileon.es (C.B.); lsanv@unileon.es (L.S.-V.)
[2] Centro de Tecnología Biomédica/Escuela Técnica Superior de Ingenieros Informáticos, Universidad Politécnica de Madrid, 28660 Madrid, Spain; alejandro.rg@upm.es
[3] SECOMUCI Research Groups, Escuela de Ingenierías Industrial e Informática, Universidad de León, Campus de Vegazana s/n, C.P. 24071 León, Spain
* Correspondence: isaias.garcia@unileon.es; Tel.: +34-98-7291000 (ext. 5289)

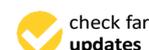

**Abstract:** Social Network Analysis (SNA) is a set of techniques developed in the field of social and behavioral sciences research, in order to characterize and study the social relationships that are established among a set of individuals. When building a social network for performing an SNA analysis, an initial process of data gathering is achieved in order to extract the characteristics of the individuals and their relationships. This is usually done by completing a questionnaire containing different types of questions that will be later used to obtain the SNA measures needed to perform the study. There are, then, a great number of different possible network-generating questions and also many possibilities for mapping the responses to the corresponding characteristics and relationships. Many variations may be introduced into these questions (the way they are posed, the weights given to each of the responses, etc.) that may have an effect on the resulting networks. All these different variations are difficult to achieve manually, because the process is time-consuming and error-prone. The tool described in this paper uses semantic knowledge representation techniques in order to facilitate this kind of sensitivity studies. The base of the tool is a conceptual structure, called "ontology" that is able to represent the different concepts and their definitions. The tool is compared to other similar ones, and the advantages of the approach are highlighted, giving some particular examples from an ongoing SNA study about alcohol consumption habits in adolescents.

## 1. Introduction

Social Network Analysis (SNA) is a powerful technique developed in the field of social and behavioral sciences research in order to characterize and study how the social relationships among a given group of nodes (usually people) establish and evolve, and how these relationships and their characteristics may serve as a guide for studying some existing social phenomena (homophily, influence, selection, etc.) or predicting their evolution in the future [1].

SNA techniques were first applied to organizations in order to study human groups including their economic, political or affective linkages [1]; but they soon found application in broader domains, such as health, where they have found an increasing number of uses during the last two decades [2–4].

Nursing is one of the main areas of application of SNA within the healthcare domain, and also one of the more promising ones because of the special focus of this discipline on sociodemographic issues, as is the case in epidemiology studies, healthcare education or intervention design, execution and monitoring [5].

In the context of studies devoted to substance consumption, several works have applied SNA techniques to gain insight into the effects of peers on the consumption levels of the individuals. These studies are especially important when the population object of study consists of adolescents. In the context of alcohol consumption in adolescents, a number of studies have been carried out using SNA techniques and tools [6–11]. Some studies have proven the existence of an influence mechanism that makes the individual adapt his drinking habits to those of his close friends [7,12–14].

The usual approach when using SNA to studying substance consumption in adolescents is: (1) choosing the social environment/s where the adolescent is going to be studied (the classroom, friends, family, etc.); (2) obtaining a convenient amount of information about the adolescent's characteristics and the connections he has within these different social networks in which he is immersed; and (3) performing a study using SNA techniques, that can be of two main types [15]:

- **Structural SNA Analysis:** This type of analysis is used to evaluate the situation at a given point in time and measure, for example, the consumption risk level or a given individual's role in the network.
- **Dynamics SNA Analysis:** This type of analysis is typical of longitudinal studies, usually related to a healthcare intervention program, for example, with the aim of reducing the alcohol consumption levels or disseminating some healthcare education program in a network. In this case, several information-gathering stages are performed at different points in time, with the aim of studying the evolution of the network ties or consumption habits, or finding explanations for the habit changes found in the individuals.

Finally, once the SNA analysis has been performed, the visualization and study of the resulting graphics as the 4th usual approach is the last, though very important, phase of SNA studies. Here, researchers can gain insight into the network structure, as well as the participating individuals and their positions within the network, and the results of the analysis. By visualizing different combinations of relationships and individual characteristics, researchers can build a new hypothesis that the visualization may suggest, which can lead to new research questions that can be later studied by building a different visualization of the data, or even performing a new study with subtle different attributes or characteristics.

As can be seen, the greater the flexibility of the visualization application, the better the results and information obtained. This flexibility is even more beneficial if it can be extended not only to the visualization phase, but to all the stages of the analysis. For example, many times, it is useful to test how the results of the analysis are affected by taking different initial assumptions regarding the responses of the users to the questionnaire, considering different definitions for the individual characteristics, or using different SNA measures to draw the conclusions from the study. Performing this kind of sensitivity and comparative studies is not an easy task with the SNA tools available today. Many of them cannot introduce such variations in the variables of the study, others are not able to account for, and compare, different results obtained, and no one offers an integrated environment for automating the study from the data-gathering process to the visualization of the results obtained. This leads to the need for a lot of manual processing of data at different stages, which makes the studies subjected to errors and very time-consuming.

The research described in this paper is an effort to build a conceptual architecture and a computer application devoted to performing SNA studies that solve the mentioned gaps found in the current SNA tools. With such an application, more in-depth and more complete SNA studies can be conducted, obtaining a higher amount of information about individuals with fewer errors and much faster than when using the usual approaches. The work has been carried out in the context of an ongoing study program about alcohol consumption habits of adolescents, where their social connections, established in their educational environments with their classroom and school peers, are the main social variable to be studied [16]. An initial application was built in order to automate many of the tasks involved in an SNA study. This paper describes the evolution of this tool in order to add the flexibility described in the previous paragraph.

The rest of this paper is organized as follows. Section 2 describes the ongoing study about alcohol consumption in adolescents and the first application developed in order to automate all of the phases of the SNA study. Two example scenarios are presented in which the extra flexibility in these SNA studies would be useful. Section 3 describes the tool that has been implemented to add these new features; the underlying knowledge model (an ontology) and its use are presented and exemplified. Section 4 gives some discussion and compares the functionalities and the approach of the tool to other similar approaches, stressing the advantages of the one described in this paper. Finally, Section 5 presents some conclusions and future work in order to extend and improve the functionality of this kind of applications.

## 2. Initial Application for Automating SNA Studies on Alcohol Consumption in Adolescents

As was stated previously, the studies carried out by using SNA techniques involve a number of data collection, curation and analysis phases that are usually carried out manually, giving rise to errors and being very time-consuming. These issues limit the results that could be obtained from the analysis and also the scope of many studies. Having this fact in mind, a tool that is able to automate all the phases of the SNA study was designed and constructed, as described in [17]. The tool presented in this paper is an evolution of the previous one, aimed at gaining flexibility and features, and allowing obtaining insight into the social network object of study with a number of new functionalities based on the semantic representation of the information being managed. In this section, the initial application is built, and some examples of the issues that give rise to the need for new features are presented.

### 2.1. Initial Application for the Automation of SNA Studies

The kind of functionality and flexibility that is required from an application able to automate the process behind an SNA study can be achieved by using knowledge modeling techniques to represent the data in a computer. Building a semantic knowledge model to represent the data and the analysis processes involved in the SNA, applied to alcohol consumption, involves the work of experts from the fields of study (experts in SNA and healthcare domains) and knowledge engineers (experts in building knowledge models). A continuous collaborative work is needed in order to obtain the expert knowledge involved in this kind of studies.

The result of this work is a conceptual structure called "ontology", which is a computer-processable representation of this expert knowledge [18]. An ontology consists of a hierarchy of concepts that represent the terms used in the domain being modeled, the relationships that hold between them, and a series of logic clauses and/or rules for defining new concepts, properties or relationships. For example, in this case, it states that a questionnaire is composed of questions, and questions may be of different types. Each type of question has its own knowledge model (the text of the question and the kind of possible answers that may be obtained, the value corresponding to each of the question responses, if any, etc.). Individuals are also represented in the ontology, and a relationship holds between an individual and a questionnaire at a given time. This relationship is built from the set of answers that a particular individual has given to a particular questionnaire at a given time.

Moreover, from the responses of the individuals to questions related to their social connections, a new conceptual structure can be built: the social network composed of individuals and their relationships. From this social network, a number of SNA metrics can be obtained that have to be conceptualized in the ontology. All these concepts, and the corresponding values inserted when populating the ontology with particular data can be further processed by applying logic clauses and/or rules that can be stated as part of the conceptual model. For example: a *binge drinker* can be defined as "an individual who has six or more drinks in a single occasion" (according to his response to a question in the questionnaire); or a *popular individual* can be defined as "that one who has been named by four or more peers as a friend".

The power of an ontology lies in the high abstraction level used for the definition of the concepts and their relationships. In contrast to a database, where some of the relationships between concepts are missing or hardcoded in the Structured Query Language (SQL) sentences or stored procedures, all the definitions in an ontology are clearly stated and separated from any processing code that may use it.

The ontology is a standalone conceptual structure, and so the next phase, once the ontology is built, is the construction of an application that uses this knowledge model and exposes a graphical interface to the users. The last stage is a validation and evaluation process that helps to assure that the representation is correct and the system behaves as the user needs. An initial knowledge model and the corresponding application were built and used by healthcare professionals [17]. This application is able to automate the data gathering, curation and visualization processes of the main SNA measures corresponding to a given social network and their participants.

The application allows the creation of a questionnaire for data gathering. This questionnaire included information about the adolescent personal data (age, gender, address, etc.), his auto perceived welfare status, his family affluence level, his alcohol consumption habits, etc. The questionnaire also included two questions for generating a pair of social networks: one to study the friendship relationships and another to determine drinking mates. The following pre-existing, validated questionnaires were also included:

- Alcohol Use Disorders Identification Test (AUDIT) [19]. This test is used to detect problematic levels of alcohol consumption, or dependence.
- Family Affluence Scale II (FAS II) [20]: used for assessing family wealth.
- ESTUDES (Encuesta Sobre Uso de Drogas en Enseñanzas Secundarias en España—poll about drug use in secondary school in Spain-) [21]: a biannual study for gaining insight into behaviors and attitudes about substance use. It is a test that includes questions about consumption of different substances. Those related to alcohol were discarded because they were obtained in other parts of the main questionnaire.
- KIDSCREEN-27 (health-related quality of life questionnaire for children and young people and their parents) [22]: used for knowing the quality of life perception of the adolescent based on five scales—physical and psychological well-being, autonomy and parent relation, peers and social support and school environment.
- Self-efficacy [23] (Spanish adapted version): assessing the belief of the adolescent about his/her own capacities to achieve different goals, specially facing stressful situations.

By including the questionnaire as part of the knowledge model, the process of data curation can be automated, without the need of the usual spreadsheet manipulations or manual matrices generation processes. Further details of the questionnaire conceptualization are given in [17]. The data are formatted into suitable representations in order to be fed to the analysis module that will generate the social networks and present the corresponding SNA measures, as well as the information obtained for each of the individuals (the application includes an anonymization feature that preserves the identity of the participants). As the knowledge model includes the conceptualization of the questionnaire, any social network that needs to be studied will only need to have the corresponding network-generating question introduced in the questionnaire.

Figure 1 shows a screenshot of the application, showing the friendship relationship network for a given classroom, while Figure 2 shows the information about a given student in the classroom, including the information obtained from the questionnaire and from the SNA analysis performed.

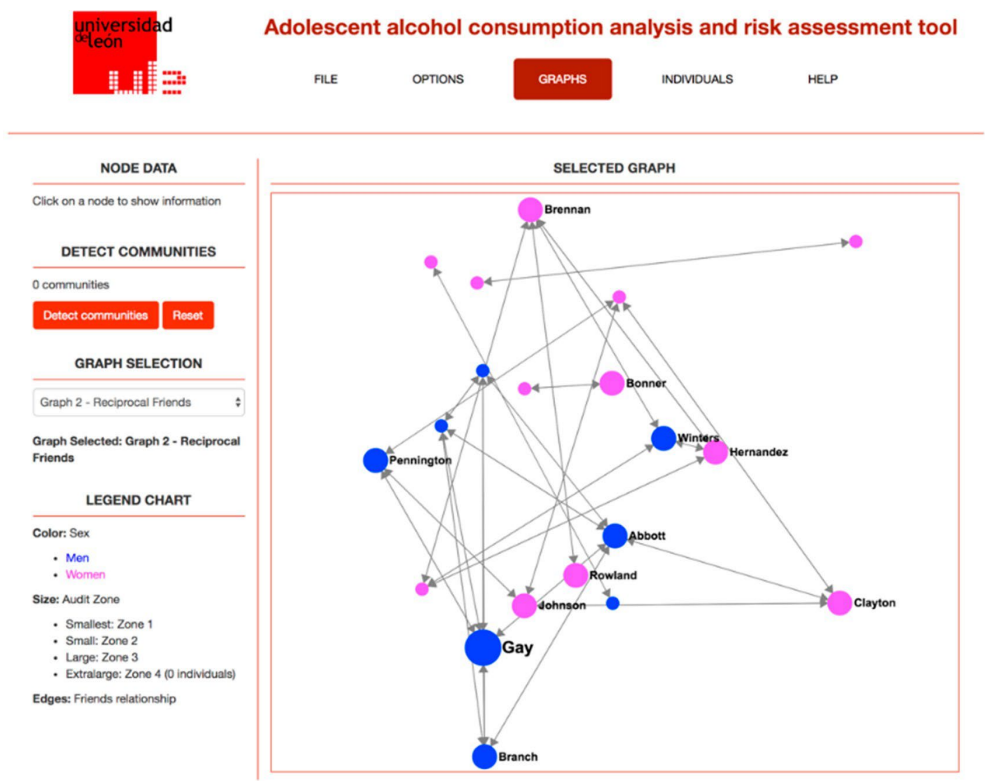

**Figure 1.** Friendship network in a classroom.

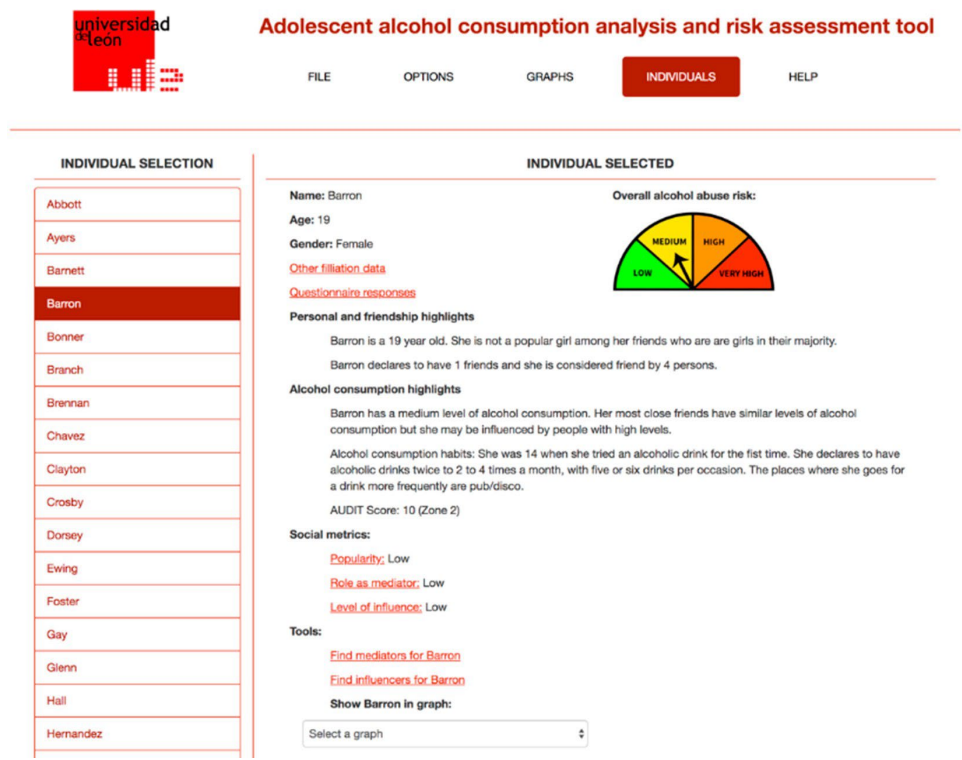

**Figure 2.** Information about an individual including the results of the SNA analysis. SNA: Social Network Analysis.

*2.2. The New Functionalities for Sensitivity Analysis*

Once the application described in the previous section was built and used, a number of issues and concerns about the different conclusions obtained with the information provided by the questionnaires appeared. The application performed calculations of the individual's characteristics and his social relationships by applying the formulae and reasoning processes provided by the SNA techniques and the healthcare experts. However, this kind of studies may yield slightly, or even very, different results, depending on how some of the relationships and characteristics of the individuals are defined or interpreted. For example, regarding the alcohol consumption habits, how must "binge drinking" (heavy episodic drinking) be defined? What are the variables, and their corresponding values, which must be considered when measuring it? Additionally, how will a different definition (for example, by introducing small changes in the values of the variables) affect the conclusions drawn from the study? The same can be said of the characterization of the social relationships that build, for example, the concept of "friendship"; how should this concept be defined and evaluated? What are the measured values, or combination of values, which define a friendship relationship? How many levels of friendship should be considered? Moreover, what are the consequences for the conclusions obtained when using different definitions for the concept of friendship?

Having these considerations in mind, a new set of functionalities was proposed for performing such studies within the initial tool. This resulted in an extension of the initial computer application aimed at performing these types of sensitivity analyses. The examples mentioned in the previous paragraph are presented and introduced in the next section in order to clarify the problem to be faced.

*2.3. Example Case 1: Defining and Studying a Relationship from Collected Data: the Case of Friendship*

Friendship is a complex and dynamic social relationship whose definition is known to change over time. There is a difference between childhood, adolescence and adulthood when talking about the main components of a friendship relationship. In adolescence, the sharing of activities (mainly during free time) is known to have a greater impact on friendship than other considerations [24].

When using SNA to study friendship, a network-generating question is needed in order to capture these social ties. Some approaches leave the choice to the individual by asking him to directly name his friends. Other approaches use a generating question where the individual is asked how many activities or how much time he spends with each of his peers in the classroom.

Even when choosing a given approach for the definition of friendship, the network-generating question will usually have different measures of strength for the relationship. Within the study where this research was developed, different works were surveyed in order to find the most suitable generating question. As a result of this revision process, the following question was used to generate the friendship network:

*"How much time do you spend with each of the following classmates?"*

Adolescents were able to choose the degree of contact with each of their peers by choosing one of five possible statements. Each of these statements has an associated weight (see Table 1) that is later used to quantify the relationship.

**Table 1.** Possible answers and their corresponding weights for the friendship network-generating question.

| Description/Degree of Contact | Weight |
|---|---|
| We never spend time together. | 1 |
| We sometimes spend time together. | 2 |
| We spend quite a lot of time together. | 3 |
| We are almost always together. | 4 |
| We are always together. | 5 |

After obtaining the data, the first decision is to decide how to define the existence and the degree of a friendship relationship. Should it be considered as a weighted relationship (preserving different levels of friendship during the analysis phase)? or as a dichotomous variable (does friendship exists or not?)? A usual decision, in order to keep things simple and easier for later analysis, is the second choice. However, there are still some considerations to be taken into account once this decision is taken. For example, how can it be evaluated whether a friendship relationship exists between two given individuals from the data provided by the generating question?

Moreover, friendship is said to need reciprocity in order to be recognized between two persons (that is, both individuals must have mentioned the other one in order to consider that a friendship relationship exists). However, then another important decision must be made: what should be the minimum strength (weight, as it appears in Table 1) of the links for considering the existence of friendship? Should a minimum be mandatory for each of the links? Should a minimum be required regarding the sum of the weights, or regarding the mean value? Should both considerations be taken into account? As can be seen, there are a great number of possibilities, and choosing one or another will yield a different set of results from the SNA analysis. This is the kind of sensitivity study that has to be possible with the new functionalities implemented in the computer application.

*2.4. Example Case 2: Defining a Qualitative Characteristic of an Individual: the Concept of "Binge Drinker"*

"Binge drinking" is a concept introduced in the 1990s to describe episodic high levels of alcohol consumption [25]. Many studies consider that binge drinking during youth has many significant consequences, including physiological changes in parameters, such as blood pressure and state anxiety immediately following consumption, and cognitive impairment to memory, attention and execution [26].

Initially, the concept of "binge drinking" was used to refer to a consumption pattern of a given amount of alcohol within a single occasion. Since the introduction of the concept, several definitions and interpretations have been used, including considering both different quantities of alcohol consumed and the period of time over which the consumption occurs. While some authors use the number of drinks consumed, other definitions are based on the blood alcohol concentration (BAC) resulting from the consumption pattern. This way, the usual measure for considering binge drinking is a BAC of 0.08 gram percent or above. This concentration roughly corresponds to an intake of 5 or more alcoholic drinks in men and 4 or more in women. However, this measure may drastically affect the amount of alcohol consumed, depending on the type of drink considered and, more importantly, on the serving size of the corresponding drink (that may vary substantially depending on the country). Regarding the period of time considered for the intake, the usual period was initially "an occasion" while the current use only considers two hours. A more detailed historical overview of this "binge drinking" concept can be consulted in [27]. Finally, making the concept more confusing, some authors use other names with subtle differences for this heavy alcohol intake, for example, heavy episodic drinking, risky single-occasion drinking, heavy sessional drinking, heavy drinking and risk drinking.

Using the concept of "binge drinking", which refers to (and characterizes) an alcohol consumption episode, a new characterization may be defined for a given individual who presents an alcohol consumption pattern that includes such episodes of "binge drinking". An individual who has a given number of "binge drinking" episodes in a given time may be considered a "binge drinker" or, as is also known, a "heavy alcohol user". This new characterization, which applies to a given individual instead of an alcohol consumption episode, may be very valuable if it is included as part of the SNA study. However, as is the case with "binge drinking", the concept of "binge drinker" or "heavy alcohol user" has a number of different definitions. One example would be "that individual who binges drink for five or more times during a month", but different definitions, even different levels of "binge drinker", exist.

Finally, it is also worth noting that, often, it is not possible to obtain a detailed description of the alcohol intake as is desirable and necessary for an accurate calculation of the "binge drinker" condition (whatever the definition used) as previously defined. Instead, on such occasions, other indirect measures are used, for example, using the results of the AUDIT to obtain an approximation of the concept. As will be presented next, this approach introduces new uncertainties in the definition of the concept that must also be taken into account for the analysis phase.

## 3. Results

In the previous section, two examples were presented in order to exemplify the kind of new functionalities that were desired for the SNA application described in Section 2. These functionalities are aimed at helping the development of sensitivity analyses that can yield valuable information about how small (or big) variations in characteristic calculations or relationship definitions affect the results obtained. In this section, the solution given in the current research is presented, using the two example cases to illustrate it.

*3.1. Conceptualizing and Managing the Friendship Relationship in the Computer*

By using a knowledge model, different networks can be considered for different definitions of the same relationship. In this example, the friendship relationship can be defined under different conditions established based on the questionnaire responses to the generating question. As the questionnaire is formalized in the ontology, the process can be automated by implementing a new application window where these different conditions can be stated. For example, Figure 3 shows the definition of a relationship named "strong relationship"; it is defined as a reciprocal relation where both individuals have named each other and have responded regarding their degree of contact with a statement having a weight of 4 or 5 (see Table 1).

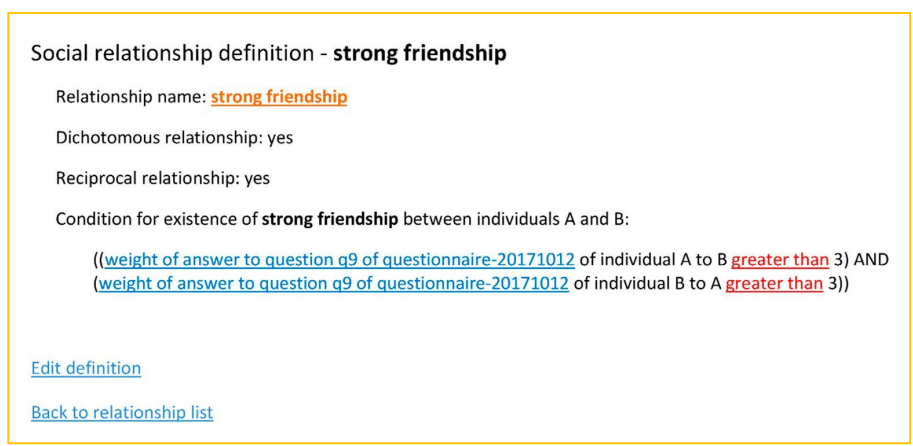

**Figure 3.** Screenshot showing the definition of a relationship.

Once these conditions are established, a name is given to the corresponding social network that is generated from it, and this network will be automatically calculated and stored in the ontology for later use, using a logic-based formalism in order to be processable by the computer. The user can generate and compare as many networks as he desires, combining the network definitions with different characteristic calculations in order to subsequently perform a number of sensitivity analyses.

Figure 4 shows different social networks and the condition that has been used for each case. By using a fixed arrangement of nodes, a visual comparison is easy to achieve by simply choosing the different tabs where the networks are drawn. As well as visually, a different set of SNA measures can also be automatically obtained and studied in the application. By comparing the results obtained by different sets of friendship definitions, the expert can evaluate how these results differ, helping him to choose a given definition to be used for subsequent analyses, or being warned about how the definitions affect the interpretation of the results.

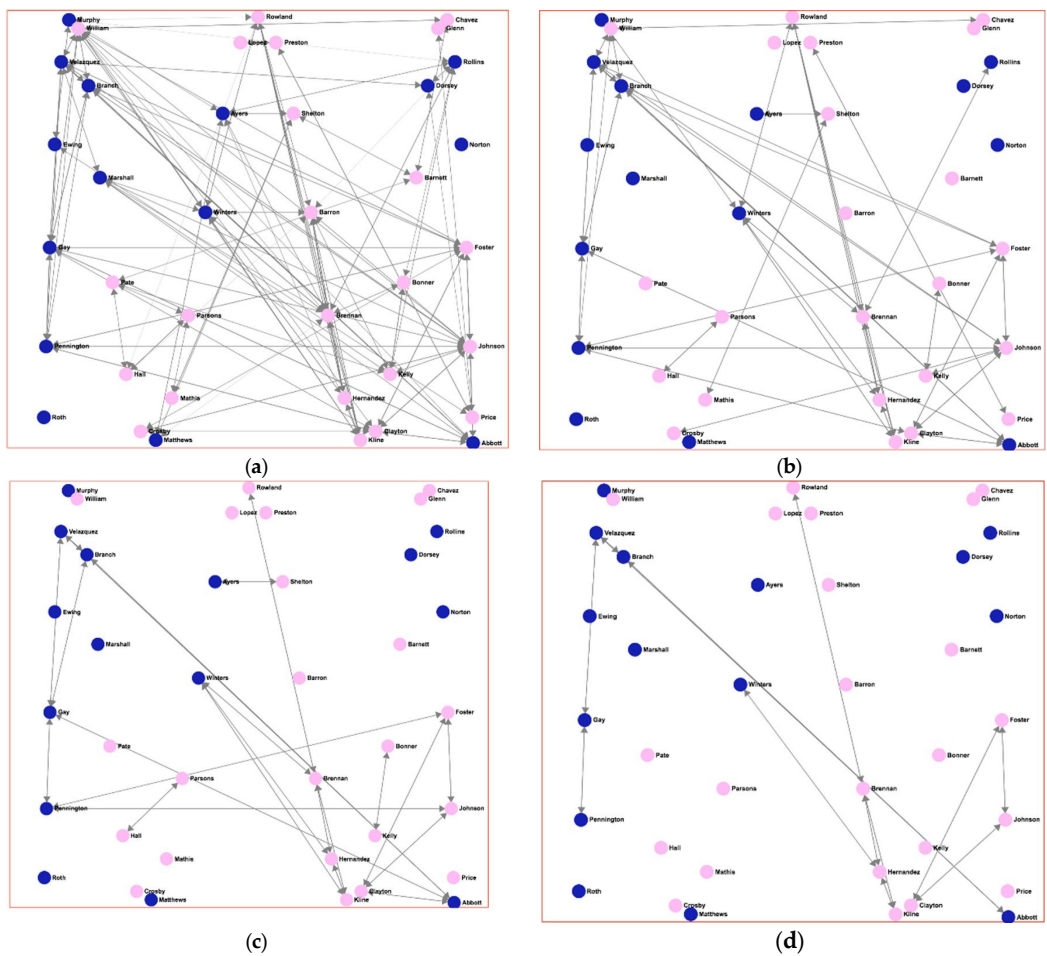

**Figure 4.** Different friendship definitions give rise to different networks. Pink (light grey) nodes represent women, and dark blue (dark grey) nodes represent men. Names are anonymized. (**a**) Resulting network considering friendship as reciprocal mentions with weights of >1; (**b**) resulting network considering friendship as reciprocal mentions with weights of >2; (**c**) resulting network considering friendship as reciprocal mentions with weights of >3 ("strong friendship"); (**d**) resulting network considering friendship as reciprocal mentions with weights of >4 ("best friends").

### 3.2. Conceptualizing the Concept of "Binge Drinker"

As seen in the previous section, the concepts of "binge drinking" and "binge drinker" have no fixed definitions. A great variation can be found from one author, or institution, to another, and this variation is even remarkable from one country to another, because there are different measures for an "average drink" and different definitions of the very concept of "binge drinking".

Alcohol consumption, the questionnaire used for data gathering included the AUDIT. This test is a tool for obtaining information regarding consumption and characterization of the consumption risk level of the individual according to his responses to 10 different questions (see Table 2).

**Table 2.** Alcohol Use Disorders Identification Test (AUDIT) and the corresponding scores for each question.

| AUDIT | Score for Each Response | | | | |
|---|---|---|---|---|---|
| | 0 | 1 | 2 | 3 | 4 |
| 1. How often do you have a drink containing alcohol? | Never | Monthly or less | 2–4 times a month | 2–3 times a week | 4 or more times a week |
| 2. How many drinks containing alcohol do you have on a typical day when you are drinking? | 0–2 | 3 or 4 | 5 or 6 | 7–9 | 10 or more |
| 3. How often do you have four or more drinks on one occasion? | Never | Less than monthly | Monthly | Weekly | Daily or almost daily |
| 4. How often during the last year have you found that you were not able to stop drinking once you had started? | Never | Less than monthly | Monthly | Weekly | Daily or almost daily |
| 5. How often during the last year have you failed to do what was normally expected of you because of drinking? | Never | Less than monthly | Monthly | Weekly | Daily or almost daily |
| 6. How often during the last year have you needed a first drink in the morning to get yourself going after a heavy drinking session? | Never | Less than monthly | Monthly | Weekly | Daily or almost daily |
| 7. How often during the last year have you had a feeling of guilt or remorse after drinking? | Never | Less than monthly | Monthly | Weekly | Daily or almost daily |
| 8. How often during the last year have you been unable to remember what happened the night before because of your drinking? | Never | Less than monthly | Monthly | Weekly | Daily or almost daily |
| 9. Have you or someone else been injured because of your drinking? | No | | Yes, but not in the last year | | Yes, in the last year |
| 10. Has a relative, friend, doctor, or other health care worker been concerned about your drinking or suggested you cut down? | No | | Yes, but not in the last year | | Yes, in the last year |

The questionnaire to be completed by the students was rather long (it already included a number of questions for different purposes, such as family affluence levels, drinking mates, other substance consumptions, welfare auto-perception, etc.), and therefore, it was decided not to complicate it further with some new questions to accurately calculate the binge drinking behavior (as stated in Section 2.4). Instead, it was decided to also use the AUDIT as a tool to characterize heavy alcohol consumers and binge drinkers, because some studies have demonstrated this possibility [7,28–30].

The previously mentioned studies propose quite different cut-off values for classifying an individual as a "binge drinker", based on the score obtained from the first three questions of the AUDIT (see Table 2). For example, in a study carried out in Finland [28], the optimal cut-off points for males were found to be greater than or equal to 7 or 8. Among females, the optimal cut-off points were found to be greater than or equal to 5. In a study carried out among Spanish students [29], the best cut-off point was found to be 4, regardless of gender. The choice of the cut-off value is always a compromise between sensitivity (the ability to include the majority of the real binge drinkers in the interval) and specificity (the ability to correctly identify non-binge drinkers, that is, not classifying many non-binge drinkers in the binge-drinker category because of the cut-off value).

As can be seen, using the AUDIT to characterize an individual as a "binge drinker" is controversial; nevertheless, what must be stressed here is that, using a semantic definition of the way the characterization has been defined, the analyst can take this fact into account in order to interpret the results yielded. Moreover, the same study can be achieved using different definitions of the concept, and the results can be compared in order to see the differences and the suitability of each one of them.

Figure 5 shows the "strong friendship" network of Figure 4c. In this case, squared nodes, rather than rounded ones, indicate the individuals considered as "binge drinkers". Pink-filled nodes (light grey) represent girls, while dark blue ones (dark grey) represent boys. In network (5a), individuals are characterized as "binge drinkers" using the criteria of the AUDIT score greater than 7 for males or greater than 4 for females; definition taken from [28], the Finnish study. In network (5b), individuals are characterized as "binge drinkers" when the AUDIT score is greater than 4, regardless of gender, definition taken from [29], a Spanish study.

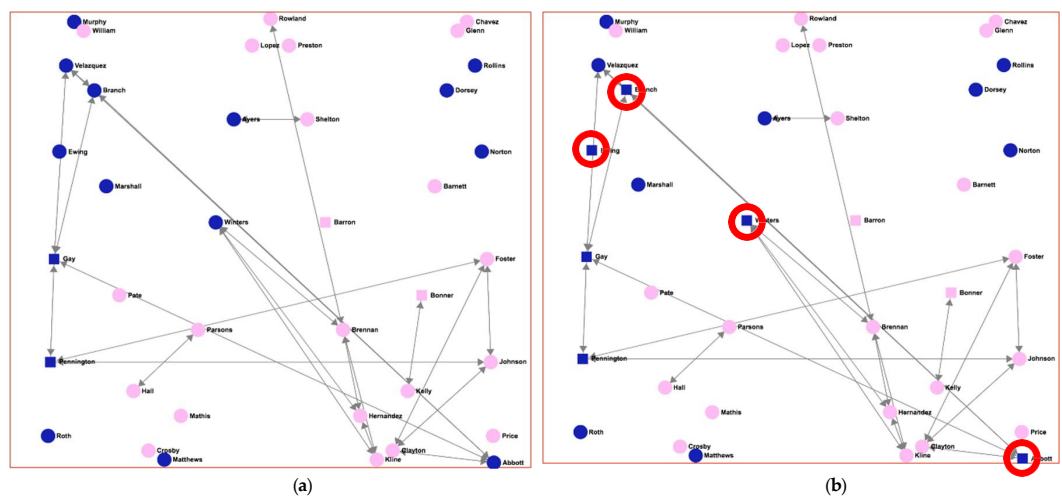

**Figure 5.** The same "strong friendship" network, but showing binge drinkers indicated by squared nodes. (**a**) Binge drinkers using the definition in Aalto et al. [28]; (**b**) binge drinkers using the definition in Tomás et al. [29].

As can be seen, there is a great difference between the two graphs regarding the number of binge drinkers in the network. In Figure 5b, there are four extra males that are now considered as binge drinkers (marked with red circles), which gives a total of six male binge-drinker individuals, compared to those in Figure 5a where only two males were considered binge drinkers. This difference would result in very different analysis results if the characterization of "binge drinker" were used for later reasoning about the individuals, and the final outcomes of the study would also be very different. For example, assuming (as many researches have pointed out) that there is an influence pattern regarding alcohol consumption in adolescents, an individual who is a binge drinker and has a great number of friends (high value of the *degree* SNA measure) could be qualified as a "bad influence" for the individuals who are his friends. Furthermore, this qualification could lead to further new consequences and reasoning (for example, initiating an intervention on a given set of individuals). (Please confirm whether the italic is necessary.)

Using a knowledge model, the user can know, at any time, what is the definition of "binge drinker" that is being used and also compare the results obtained by using other different definitions. Moreover, as the analysis may be completely automated in the computer, this comparison could also be achieved automatically. The user can, then, perform an initial analysis in order to decide which of the concept's definitions is more suitable for his purpose, or use a number of them and present the comparative results.

*3.3. Testing the Application*

The application based on this semantic knowledge representation approach was used to perform a study on adolescent's alcohol consumption habits in nine classrooms across three different secondary schools in Spain. The data gathering questionnaire was completed by 214 students. A total of 145,520 question items were obtained and used for drawing conclusions about the consumption patterns observed. Different sensitivity analyses were carried out using different definitions for the friendship relationship, the binge drinker concept, and the family affluence level.

**4. Discussion**

As was seen in the previous examples, many of the relationships or characteristics that are involved in an SNA study have no precise definition, because they are based on social conventions or calculations of variables created and described in different studies. As a result, the outcomes of an SNA analysis may yield quite different results when using these different definitions, and therefore, a tool able to manage this variability is very useful. In the case of the study of alcohol consumption in adolescents, the concept of friendship is one of the main relationships involved that has different possible definitions. Additionally, characteristics related to binge drinking, family affluence levels, welfare status or substance consumption levels can be defined and graded very differently. Using one definition or another for each of these concepts can lead to very diverse results.

Combining $n$ definitions of friendship with $m$ binge drinking definitions, only to show the two examples introduced, a total of $n \times m$ different SNA analyses can be conducted. Conducting these analyses without an automatic tool would be very time-consuming. If more than two concepts were involved, the number of combinations, and the time spent would rise geometrically.

Using knowledge modeling techniques to represent these different definitions, the analyses can be automated, and the results can even be automatically compared by using statistical techniques. Using knowledge modeling techniques, each of the definitions of the relationships and characteristics can be stated in a formal, logic-based structure and stored in an ontology for later use. The concepts need to be defined only once and then can be used at any time, even for defining new concepts. Moreover, as the knowledge model in the ontology is independent of the particular application, it could be used in other different applications.

Some SNA tools implement similar functionalities to these described in this paper, to a greater or lesser extent. Meerkat is a tool for visualization and community mining of social networks [31]. It uses an approach based on the concepts of filtering and extraction in order to obtain sub-graphs, which can be later used to study some specific social sub-networks where only some individuals are involved. Attributes for nodes or edges, captured in the data-gathering process or calculated, are used to establish filters: "A researcher may define a filter on any user-defined attributes or on a selection of pre-defined attributes, such as node degree and edge weight. The filter creation interface allows for a series of nested conjunctive and disjunctive Boolean operators. Furthermore, multiple filters may be defined and then enabled or disabled to create either a complex or simple query, without needing to redefine everything twice. When a network has been filtered to include only a subset of entities, or if the user has manually selected entities, these nodes can be extracted to another graph. Meerkat will open a new visualization tab, with only those nodes that passed the filter or were manually selected. This allows an analyst to make small or large changes to a graph, and flip back while comparing any differences in statistics, in node groupings for deterministic layouts, or in the communities discovered by mining algorithms".

Netminer (Cyram company, Seognam, South Korea) [32] is a software suite aimed at network exploratory analysis, including networks with large sets of data. It is based on an underlying programming platform that can be extended and used by means of scripts. "What-if analysis" of NetMiner "enables an intuitive mouse action on the visualized result for testing researcher's hypothesis during the interpretation process　　For example, users can check the changed community structure instantly when some links are dropped or can investigate the changed rank of network centrality when a node of highest betweenness centrality is deleted. For this purpose, the result can be modified by giving changes on a visualized map and numerical variables in NetMiner".

As can be seen, both Meerkat and Netminer tools allow for the implementation of what-if (or sensitivity) analyses, but using slightly different approaches. The first one uses different filters in order to generate sub-networks, of which nodes accomplish some kind of condition established over the node characteristics. Netminer uses a powerful graphical user interface in order to interact with the resulting network to modify or extract sub-networks that can be used for later analysis. Both tools use their own internal data structures in order to store the application information.

The approach used in this research allows for the definitions of not only different networks based on different relationship calculations (as is the case of the friendship relationship), but also the use of different definitions for the characteristics of the individuals (as is the case of "binge drinker"). Moreover, these different definitions are stored in an ontology that is integrated with the rest of the data involved in the SNA analysis from the data gathering with questionnaires to the results of the analysis process. By using this kind of knowledge representation in the computer, the user can make use of the different definitions at any point in the analysis process, and even combine them in order to perform complex reasoning processes and comparative studies. The ontology is an independent conceptual structure; it is not part of the application, but accessed by it by means of an application program interface (API).

It is known that different study scenarios may be best explained by different measures. For example, it is not the same to study the friendship relationship regarding alcohol consumption, academic performance, popularity, etc. According to Valente [2], "New measures for newly created concepts may be needed as increased understanding and further exploration of networks evolve". The semantic representation of characteristics proposed in this tool can be used to generate and evaluate these new measures and incorporate them into the analysis, testing their suitability for the objectives of the study.

## 5. Conclusions

This work stresses the need and shows the advantages of knowledge-based computer tools that are able to meet the needs of healthcare professionals and researchers when facing studies involving social networks, as is the case of substance consumption habits in adolescents. Many parameters used to characterize individuals in a social network study do not have a clear definition, and therefore, this fact must be taken into account when performing this kind of studies. Additionally, the nature of the social ties that build concepts such as "friendship" is not unequivocally established and varies between researchers and approaches. A tool able to capture, use and compare the different results obtained with different approaches to these calculations and definitions is highly valued. The tool presented in this paper has achieved this objective by creating and using a conceptual knowledge model (an ontology) able to reflect and account for these variabilities in the definition of the concepts.

This ontology-based approach also has some important issues to be taken into account, in order to make it very useful for building complex, useful tools. Using erroneous definitions or making mistakes when introducing new concepts in the ontology would affect the results of all the analyses using such constructs, and therefore, the process of building the ontology must be very careful. Additionally, explanation capabilities should be built into these tools, especially for complex concept definitions or reasoning processes, in order to know exactly what definitions and processes have been used to obtain the results.

Future work using the idea introduced in this research includes the use of this ontology-based knowledge representation strategy for building tools devoted to performing dynamical network analysis, longitudinal studies (such as the ones found when performing an intervention on a group of adolescents), complex hypothesis testing studies, etc.

**Author Contributions:** J.A.B.-A. and A.R.-G. designed the research objectives and constructed the ontology as knowledge engineering experts. J.A.B.-A. programmed the application. C.B. and L.S.-V. were the experts in the SNA and the healthcare domains who provided the knowledge to be represented in the ontology. C.B. and I.G. worked on the validation of the ontology. L.S.-V. worked as the healthcare user for testing the application. I.G. coordinated the work and wrote the paper.

**Funding:** This research was conducted with no external funding.

**Conflicts of Interest:** The authors declare that there is no conflict of interest regarding the publication of this paper.

## References

1. Wasserman, S.; Faust, K. *Social Network Analysis: Methods and Applications*, 1st ed.; Cambridge University Press: Cambridge, UK, 1994; ISBN 978-0521382694.
2. Valente, T.W. *Scoial Networks and Health: Models, Methods, and Applications*, 1st ed.; Oxford University Press: New York, NY, USA, 2010; ISBN 978-0195301014.
3. Chambers, D.; Wilson, P.; Thompson, C.; Harden, M. Social network analysis in healthcare settings: A systematic scoping review. *PLoS ONE* **2012**, *7*. [CrossRef] [PubMed]

4. Bae, S.H.; Nikolaev, A.; Young Seo, J.; Castner, J. Health care provider social network analysis: A systematic review. *Nurs. Outlook* **2015**, *63*, 566–584. [CrossRef] [PubMed]
5. Parnell, J.M.; Robinson, J.C. Social network analysis: Presenting an underused method for nursing research. *J. Adv. Nurs.* **2018**, *74*, 1310–1318. [CrossRef] [PubMed]
6. Duncan, G.; Boisjoly, J.; Kremer, M.; Levy, D.; Eccles, J. Peer effects in drug use and sex among college students. *J. Abnorm. Child Psychol.* **2005**, *33*, 375–385. [CrossRef] [PubMed]
7. Ennett, S.T.; Bauman, K.E.; Hussong, A.; Faris, R.; Foshee, V.A.; Cai, L.; DuRant, R.H. The peer context of adolescent substance use: Findings from social network analysis. *J. Res. Adolesc.* **2006**, *16*, 159–186. [CrossRef]
8. Mulassi, A.H.; Borracci, R.A.; Calderon, J.G.E.; Vinay, P.; Mulassi, M. Social networks on smoking, alcohol use and obesity among adolescents attending a school in the city of Lobos, Buenos Aires. *Arch. Argent. Pediatr.* **2012**, *110*, 474–484. [CrossRef] [PubMed]
9. Mundt, M. Social network analysis of peer effects on binge drinking among US adolescents. *Soc. Comput. Behav. Model.* **2013**, *LNCS 7812*, 123–134. [CrossRef]
10. Ali, M.M.; Amialchuk, A.; Nikaj, S. Alcohol consumption and social network ties among adolescents: Evidence from Add Health. *Addict. Behav.* **2014**, *39*, 918–922. [CrossRef] [PubMed]
11. Gommans, R.; Müller, C.M.; Stevens, G.W.J.M.; Cillessen, A.H.N.; Ter Bogt, T.F.M. Individual popularity, peer group popularity composition and adolescents' alcohol consumption. *J. Youth Adolesc.* **2017**, *46*, 1716–1726. [CrossRef] [PubMed]
12. Snijders, T.A.B.; Steglich, C.E.G.; Schweinberger, M. Modeling the co-evolution of networks and behavior. In *Longitudinal Models in the Behavioral and Related Sciences*, 1st ed.; van Montfort, K., Oud, H., Satorra, A., Eds.; Lawrence Erlbaum Associates Publishers: Mahwah, NJ, USA, 2007.
13. Moody, J.; Brynildsen, W.D.; Osgood, D.W.; Feinberg, M.E.; Gest, S. Popularity trajectories and substance use in early adolescence. *Soc. Netw.* **2011**, *33*, 101–112. [CrossRef] [PubMed]
14. Osgood, D.W.; Ragan, D.T.; Wallace, L.; Gest, S.D.; Feinberg, M.E.; Moody, J. Peers and the emergence of alcohol use: Influence and selection processes in adolescent friendship networks. *J. Res. Adolesc.* **2013**, *23*, 500–512. [CrossRef] [PubMed]
15. Benhiba, L.; Loutfi, A.; Abdou, M.; Idrissi, J.A.J. A Classification of healthcare social network analysis applications. *Biostec 2017* **2017**, 978–989. [CrossRef]
16. Quiroga, E.; Pinto-Carral, A.; García, I.; Molina, A.J.; Fernández-Villa, T.; Martín, V. The influence of adolescents' social networks on alcohol consumption: a descriptive study of Spanish adolescents using social network analysis. *Int. J. Environ. Res. Public Health* **2018**, *15*, 1795. [CrossRef] [PubMed]
17. Benítez, J.A.; Labra, J.E.; Quiroga, E.; Martín, V.; García, I.; Marqués-Sánchez, P.; Benavides, C.A. Web-based tool for automatic data collection, curation, and visualization of complex healthcare survey studies including social network analysis. *Comput. Math. Methods Med.* **2017**, *4*, 1–8. [CrossRef] [PubMed]
18. Zenuni, X.; Raufi, B.; Ismaili, F.; Ajdari, J. State of the art of semantic web for healthcare. *Procedia – Soc. Behav. Sci.* **2015**, *195*, 1990–1998. [CrossRef]
19. Babor, T.F.; Higgins-Biddle, J.C.; Saunders, J.B.; Monteiro, M.G. *The Alcohol Use Disorders Identification Test Guidelines for Use in Primary Care*, 2nd ed.; World Health Organization: Geneva, Switzerland, 2001.
20. Currie, C.E.; Elton, R.A.; Todd, J.; Platt, S. Indicators of socioeconomic status for adolescents: The WHO health behaviour in school-aged children survey. *Health Educ. Res.* **1997**, *12*, 385–397. [CrossRef] [PubMed]
21. Spanish Ministry of Health, Ministerio de Sanidad y Consumo. ESTUDES (2014): Poll about Drug Use in Secondary Schools in Spain. National Drugs Plan—ESTUDES Poll. 2014. Available online: http://www.pnsd.mscbs.gob.es/en/profesionales/sistemasInformacion/sistemaInformacion/pdf/2016_Informe_ESTUDES.pdf (accessed on 28 October 2018).
22. The KIDSCREEN Questionnaires—Quality of Life Questionnaires for Children and Adolescents. Available online: https://www.kidscreen.org/english/questionnaires/ (accessed on 28 October 2018).
23. Schwarzer, R.; Jerusalem, M. Generalized Self-Efficacy scale. In *Measures in Health Psychology: A User's Portfolio*; Weinman, N.J., Wright, S., Johnston, M., Eds.; NFER-NELSON: Slough, UK, 1995; pp. 35–37. ISBN1 0708707335. ISBN2 9780708707333.
24. Sherman, A.M.; de Vries, B.; Lansford, J.E. Friendship in childhood and adulthood: Lessons across the life span. *Int. J. Aging Hum. Dev.* **2000**, *51*, 31–51. [CrossRef] [PubMed]

25. Wechsler, H.; Isaac, N. 'Binge' drinkers at Massachusetts colleges: Prevalence, drinking style, time trends, and associated problems. *J. Am. Med. Assoc.* **1992**, 2929–2931. [CrossRef]
26. López Caneda, E.; Mota, N.; Crego, A.; Velasquez, T.; Corral, M.; Rodríguez Holguín, S.; Cadaveira, F. Neurocognitive anomalies associated with the binge drinking pattern of alcohol consumption in adolescents and young people: A review. *Adicciones* **2014**, *26*, 334–359. [CrossRef]
27. Kuntsche, E.; Kuntsche, S.; Thrul, J.; Gmel, G. Binge drinking: Health impact, prevalence, correlates and interventions. *Psychol. Heal.* **2017**, *32*, 976–1017. [CrossRef] [PubMed]
28. Aalto, M.; Alho, H.; Halme, J.T.; Seppä, K. AUDIT and its abbreviated versions in detecting heavy and binge drinking in a general population survey. *Drug Alcohol Depend.* **2009**, *103*, 25–29. [CrossRef] [PubMed]
29. Tomás, M.T.C.; Costa, J.A.G.; Motos-Sellés, P.; Beitia, M.D.S.; Mahía, F.C. The utility of the alcohol use disorders identification test (AUDIT) for the analysis of binge drinking in university students. *Psicothema* **2017**, *29*, 229–235. (In Spanish) [CrossRef]
30. Liskola, J.; Haravuori, H.; Lindberg, N.; Niemelä, S.; Karlsson, L.; Kiviruusu, O.; Marttunen, M. AUDIT and AUDIT-C as screening instruments for alcohol problem use in adolescents. *Drug Alcohol Depend.* **2018**, *188*, 266–273. [CrossRef] [PubMed]
31. Chen, J.; Fagnan, J.; Goebel, R.; Rabbany, R.; Sangi, F.; Takaffoli, M.; Verbeek, E.; Zaiane, O. Meerkat: Community Mining with Dynamic Social Networks. In Proceedings of the 2010 IEEE International Conference on Data Mining Workshops, Sydney, NSW, Australia, 13–17 December 2010.
32. Ghim, G.H.; Cho, N.; Seo, J. NetMiner. In *Encyclopedia of Social Network Analysis and Mining*; Alhajj, R., Rokne, J., Eds.; Springer: New York, NY, USA, 2014; pp. 1025–1037. ISBN 978-1-4614-6170-8.